\documentclass[aps,prl,showpacs]{revtex4}
\usepackage{graphicx}
\usepackage{amsfonts,amssymb,amsmath}
\usepackage{bm}

\begin{document}

\title{SPECTRUM OF THE RELATIVISTIC PARTICLES IN VARIOUS POTENTIALS}
\date{\today}
 
\author{Ramazan Ko\c{c}}
\email{koc@gantep.edu.tr}
\affiliation{Department of Physics, Faculty of Engineering 
University of Gaziantep,  27310 Gaziantep, Turkey}
\author{Mehmet Koca}
\email{kocam@squ.edu.om}
\affiliation{Department of Physics, College of Science,
Sultan Qaboos University, PO Box 36  \\
Al-Khod 123, Sultanete of Oman}

\begin{abstract}
We extend the notion of Dirac oscillator in two dimensions, to construct a
set of potentials. These potentials becomes exactly and quasi-exactly
solvable potentials of non-relativistic quantum mechanics when they are
transformed into a Schr\"{o}dinger-like equation. For the exactly solvable
potentials, eigenvalues are calculated and eigenfunctions are given by
confluent hypergeometric functions. It is shown that, our formulation also
leads to the study of those potentials in the framework of the
supersymmetric quantum mechanics.
\end{abstract}
\maketitle

\section{Introduction}

The solution of the $(2+1)$-dimensional Dirac equation are of special
interest, because of the rapid growth in nanofabrication technology that has
made possible to confine laterally two-dimensional (2D) electron systems.
These quantum confined electron systems are referred to as artificial atoms
where the potential of the nucleus, in the non relativistic case, is
replaced by an effective potential of the form $V=\frac{1}{2}r^{2}$ which is
often used as realistic approximation \cite{villa1,chak,villa2}. Although
the parabolic potential appears to be a good approximation for artificial
atom structures, their modelling with various potential profiles will be
interesting from the theoretical point of view as well as from its practical
applications.

In order to analyze relativistic effects on the spectrum of such physical
systems one should construct Dirac equation including adequate potentials
and obtain its solution. For the relativistic case, the spectrum and
properties of such systems can be determined by using two dimensional Dirac
oscillator\cite{villa2,schak,naagu}. Relativistic extensions of the various
exactly and quasi-exactly solvable (QES) potentials have also turned out to
be of importance in the description of 2D phenomena\cite%
{ma,lin,vaid,khal,wil}. Different condensed matter physics phenomena point
to the existence of $(2+1)$-dimensional systems whose spectrum determined by
Dirac equation Hamiltonian including various potentials. It is well known
that the dirac equation is used for the description of spin-1/2 relativistic
particle. Meanwhile we mention here that the Hamiltonian in the form of the
Klein-Gordon equation, so called Feshbach-Villars equation, has been
constructed in a two component form for spinless particles\cite{fesh} and in
an eight-component form for spin-1/2 particles\cite{merad,robson1,robson2}.
Regrettably, the Dirac equation is exactly solvable only in a very
restricted potentials. It is the purpose of the present article to construct
a Dirac equation including a class of potentials whose spectrum can exactly
be determined. For this purpose we transform the Dirac equation into two Schr%
\"{o}dinger-like equation, because there exist a large number of papers
discussing the 2D, few electron systems, most of which are tackling the
problem in the framework of the Schr\"{o}dinger-like equation.

A Dirac equation with an interaction linear in coordinates was considered
long ago\cite{ito} and recently rediscovered in the context of the
relativistic many body theories\cite{mosh1}. The equation is named \textit{%
Dirac oscillator}, since in the non-relativistic limit it becomes a harmonic
oscillator with a very strong spin-orbit coupling term. Dirac oscillator has
attracted much attention and the concept gave rise to a large number of
papers concerned with its different aspects\cite%
{moreno,mosh2,ferk,ros,ho,mart,levai,alhaidari}. Analogous to the Dirac
equation, with a modified momentum operator,which in the nonrelativistic
limit turns out be the usual Schr\"{o}dinger equation. As we have already
noted the Dirac equation including various potentials might attract much
attention because it may have some physical applications, particularly in
the condensed matter physics. It seems that one can present more realistic
models for the artificial atoms using the procedure given here.

In order to solve Dirac equation, in this paper, we use functional approach
which have been applied to solve Schr\"{o}dinger equation for a exactly or
QES potential profile. For a QES potential it is possible to determine
algebraically a part of spectrum but not whole spectrum \cite%
{turb,bender1,bender2}. Our approach also gives a hint to the solution of
the problem in the framework of the SUSYQM.

The method presented here consist of the followings. In section 2, we
introduce $(2+1)$-dimensional Dirac equation. Using the structures worked
for the Dirac oscillator we develop a method to construct a class of exactly
and QES potential profile. We transform the Dirac equation into the form of
the Schr\"{o}dinger equation. In section 3, we construct a Dirac equation
including, harmonic oscillator, Coulomb and Morse potentials. We obtain the
corresponding eigenvalues and eigenfunctions. In section 4 a class of QES
potentials are constructed and their ground state wave functions are
explicitly determined. We conclude our results in section 5.

\section{Method}

The $(2+1)-$dimensional Dirac equation for free particle of mass $m$ in
terms of two-component spinors $\psi ,$ can be written as%
\begin{equation}
E\psi =\left[ \sum_{i=1}^{2}c\beta \gamma _{i}p_{i}\mathbf{+}\beta mc^{2}%
\right] \psi  \label{d1}
\end{equation}%
Since we are using only two component spinors, the matrices $\beta $ and $%
\beta \gamma _{i}$ are conveniently defined in terms of the Pauli spin
matrices which satisfy the relation $\sigma _{i}\sigma _{j}=\delta
_{ij}+\varepsilon _{ijk}\sigma _{k},$ given by 
\begin{equation}
\beta \gamma _{1}=\sigma _{1};\quad \beta \gamma _{2}=\sigma _{2};\quad
\beta =\sigma _{3}.  \label{d2}
\end{equation}%
In $(2+1)-$dimensions, the momentum operator $p_{i}$ is two component
differential operator, $\mathbf{p}=-i\hbar (\partial _{x},\partial _{y})$,
for free particle$.$ In the presence of the magnetic field it is replaced by 
$\mathbf{p}\rightarrow \mathbf{p-}e\mathbf{A}$, where $\mathbf{A}$ is the
vector potential, and the 2D Dirac oscillator can be constructed by changing
the momentum $\mathbf{p}\rightarrow \mathbf{p-}im\omega \sigma _{3}r\widehat{%
\mathbf{r}}$. We are now seeking for a certain form of the momentum operator
that can be interpreted as exactly solvable Schr\"{o}dinger equation in the
non-relativistic limit. For this purpose we introduce the following momentum
operator%
\begin{equation}
\mathbf{p}\rightarrow \mathbf{p-}e\mathbf{A+}i\sigma _{3}v(r)\widehat{%
\mathbf{r}}.  \label{d3}
\end{equation}%
Then Dirac equation (\ref{d1}) with the momentum operator (\ref{d3}) takes
the form%
\begin{eqnarray}
\left[ E-\sigma _{0}mc^{2}\right] \psi &=&c\sigma _{+}\left[
p_{x}+ip_{y}-e\left( A_{x}+iA_{y}\right) -i\left( v_{x}+iv_{y}\right) \right]
\psi +  \notag \\
&&c\sigma _{-}\left[ p_{x}-ip_{y}-e\left( A_{x}-iA_{y}\right) +i\left(
v_{x}-iv_{y}\right) \right] \psi  \label{d4}
\end{eqnarray}%
where $p_{i},A_{i},$ and $v_{i}$ are the $i^{th}$ components of the
momentum, vector potential and the potential in Cartesian coordinate system,
respectively. In polar coordinate, $x=r\cos \phi ,\quad y=r\sin \phi $, with
the choices of the vector potential $A_{x}=-A(r)\sin \phi ,\quad
A_{y}=A(r)\cos \phi $ and the potential $v_{x}=v(r)\cos \phi ,\quad
v_{y}=v(r)\sin \phi $ the 2D Dirac equation (\ref{d4}) takes the form 
\begin{subequations}
\begin{eqnarray}
&&E_{-}\psi _{+}=ce^{i\phi }\left[ -i\hbar \frac{\partial }{\partial r}+%
\frac{\hbar }{r}\frac{\partial }{\partial \phi }-i(eA(r)+v(r))\right] \psi
_{-}  \label{d5a} \\
&&E_{+}\psi _{-}=ce^{-i\phi }\left[ -i\hbar \frac{\partial }{\partial r}-%
\frac{\hbar }{r}\frac{\partial }{\partial \phi }+i(eA(r)+v(r))\right] \psi
_{+}  \label{d5b}
\end{eqnarray}%
where $E_{\pm }=E\pm mc^{2}$ and $\psi _{\pm }=\psi _{\pm }(r,\phi )$ are
the upper and lower components of the spinor $\psi $. The substitution of
the wave functions 
\end{subequations}
\begin{equation}
\psi _{\pm }(r,\phi )=\frac{e^{-i\left( \ell \mp \frac{1}{2}\right) \phi }}{%
\sqrt{r}}f_{\pm }(r)  \label{d6}
\end{equation}%
leads to the following set of coupled differential equations 
\begin{subequations}
\begin{eqnarray}
E_{-}f_{+}(r) &=&-ic\left( \hbar \frac{\partial }{\partial r}+\frac{\hbar
\ell }{r}+\left( eA(r)+v(r)\right) \right) f_{-}(r)  \label{d7a} \\
E_{+}f_{-}(r) &=&-ic\left( \hbar \frac{\partial }{\partial r}-\frac{\hbar
\ell }{r}-\left( eA(r)+v(r)\right) \right) f_{+}(r).  \label{d7b}
\end{eqnarray}%
Our task is now to transform (\ref{d7a}) and (\ref{d7b}) in the form of the
Schr\"{o}dinger-like equation. Substitution of 
\end{subequations}
\begin{equation}
W(r)=\frac{\ell }{r}+\frac{\left( eA(r)+v(r)\right) }{\hbar };\quad E_{\pm
}=\pm ic\hbar \varepsilon _{\pm };\ \varepsilon ^{2}=\varepsilon
_{+}\varepsilon _{-},  \label{d8}
\end{equation}%
into (\ref{d7a}) and (\ref{d7b}), leads to the following expressions 
\begin{subequations}
\begin{eqnarray}
\varepsilon _{-}f_{+}(r) &=&\left( \frac{\partial }{\partial r}+W(r)\right)
f_{-}(r)  \label{d9a} \\
\varepsilon _{+}f_{-}(r) &=&\left( -\frac{\partial }{\partial r}+W(r)\right)
f_{+}(r).  \label{d9b}
\end{eqnarray}%
Notice that the result (\ref{d9a}) and (\ref{d9b}) which seem to hint at a
supersymmetric treatment of the Dirac equation, because the supersymmetric
operators can be expressed as $A^{\pm }=\left( \mp \frac{\partial }{\partial
r}+W(r)\right) .$ It is obvious that the functional form of the
superpotential $W(r)$ and $v(r)$ are the same except that the radial
function $\frac{\hbar \ell }{r}.$ Thus the superpotential of the
non-relativistic quantum mechanics can be recognized as the potential of the
relativistic quantum mechanics. It is obvious that the expressions (\ref{d9a}%
) and (\ref{d9b}) can be written in the form of the Schr\"{o}dinger-like
equation, by eliminating $f_{+}(r)$ and/or $f_{-}(r)$ between (\ref{d9a})
and (\ref{d9b}) provides the following expressions 
\end{subequations}
\begin{subequations}
\begin{eqnarray}
\left( -\frac{\partial ^{2}}{\partial r^{2}}-W^{2}(r)+W^{\prime
}(r)+\varepsilon ^{2}\right) f_{-}(r) &=&0  \label{d10a} \\
\left( -\frac{\partial ^{2}}{\partial r^{2}}+W^{2}(r)+W^{\prime
}(r)-\varepsilon ^{2}\right) f_{+}(r) &=&0.  \label{d10b}
\end{eqnarray}%
This is indeed an interesting result for the potentials $V_{\pm
}=W^{2}(r)\pm W^{\prime }(r)$ leading to a common spectrum, thus forming an
isospectral system.

In the following we analyze the solutions and the energy spectrum of the $%
(2+1)$-dimensional Dirac equation including various potentials.

\section{Exactly Solvable Potentials}

Appropriate choices of the superpotential $W(r)$ permits the construction of
the equations having exactly solvable potentials. In this section we
illustrate that for a large class of potentials, the Dirac equation in the
form of Schr\"{o}dinger type equations of (\ref{d10a}) and (\ref{d10b})
possess exact solutions.

\subsection{Harmonic oscillator}

Let us start with the well known problem, namely Dirac oscillator. The Dirac
oscillator can be constructed with the choices of the superpotential $W(r)=%
\frac{m}{\hbar }\omega _{T}r-\frac{\ell +1}{r}.$ Thus, the Schr\"{o}dinger
type equations (\ref{d10a}) and (\ref{d10b}) takes the form: 
\end{subequations}
\begin{subequations}
\begin{eqnarray}
\left( -\frac{\partial ^{2}}{\partial r^{2}}+\frac{\ell (\ell +1)}{r^{2}}%
+\left( \frac{m}{\hbar }\omega _{T}\right) ^{2}r^{2}+\frac{m}{\hbar }\omega
_{T}\left( 2\ell +3\right) -\varepsilon ^{2}\right) f_{-}(r) &=&0
\label{d11a} \\
\left( -\frac{\partial ^{2}}{\partial r^{2}}+\frac{(\ell +1)(\ell +2)}{r^{2}}%
+\left( \frac{m}{\hbar }\omega _{T}\right) ^{2}r^{2}-\frac{m}{\hbar }\omega
_{T}\left( 2\ell +1\right) -\varepsilon ^{2}\right) f_{+}(r) &=&0.
\label{d11b}
\end{eqnarray}%
In this case the vector potential $A(r)$ and scalar potential $v(r)$ are
given by 
\end{subequations}
\begin{equation}
A(r)=\frac{1}{2}Br,\quad v(r)=\frac{m}{\hbar }\omega r-\frac{2\ell +1}{r},
\end{equation}%
and the frequency $\omega _{T}$ can be expressed in terms of the Larmor
frequency as follows%
\begin{equation}
\omega _{T}=\omega +\omega _{L}=\omega +\frac{eB}{2m}.
\end{equation}%
In order to solve (\ref{d11a}), we change the variable $z=\frac{m}{\hbar }%
\omega _{T}r^{2}$, and introduce the wave function%
\begin{equation}
f_{-}(z)=Cz^{\frac{\ell +1}{2}}e^{-\frac{z}{2}}\text{ }g_{-}(z)
\end{equation}%
where $C$ is the normalization constant, then (\ref{d11a}) takes the form:%
\begin{equation}
\left[ z\frac{\partial ^{2}}{\partial z^{2}}+\left( \ell +\frac{3}{2}%
-z\right) +n\right] g_{-}(z)=0.  \label{ex1}
\end{equation}%
The natural number $n$ and energy satisfy the relation 
\begin{equation}
\varepsilon ^{2}=\frac{E^{2}-m^{2}c^{4}}{\hbar ^{2}c^{2}}=4n\frac{m}{\hbar }%
\omega _{T}.  \label{d12}
\end{equation}%
Note that the non-relativistic limit of the energy is obtained by setting $%
E=E_{nr}+mc^{2}$\ and considering $E_{nr}\ll mc^{2}$, we obtain the non
relativistic energy%
\begin{equation*}
E_{nr}=4n\hbar \omega _{T}.
\end{equation*}%
We now investigate the dependence of the energy on the spin. In order to
analyze spin effects,it is worth to obtain non-relativistic form of the (\ref%
{d11a}):%
\begin{equation}
\left( -\frac{\hbar ^{2}}{2m}\frac{\partial ^{2}}{\partial r^{2}}+\frac{\ell
(\ell +1)\hbar ^{2}}{2mr^{2}}+\frac{1}{2}m\omega _{T}^{2}r^{2}+\hbar \omega
_{T}\left( \ell +\frac{3}{2}\right) -E_{nr}\right) f_{-}(r)=0.
\end{equation}%
The corresponding Hamiltonian is the Hamiltonian of a harmonic oscillator
with an additional spin dependent term, $\hbar \omega _{T}\left( \ell +\frac{%
3}{2}\right) .$

To complete our analysis we turn our attention to the normalization of the
wave function. It is easy to see that the solution of (\ref{ex1}) is the
associated Laguerre polynomials, $L_{n}^{\ell +\frac{1}{2}}(z),$ then $%
f_{\_}(z)$ can be written as 
\begin{equation}
f_{-}(z)=Cz^{\frac{\ell +1}{2}}e^{-\frac{z}{2}}L_{n}^{\ell +\frac{1}{2}}(z).
\end{equation}%
The upper component, $f_{+}(z)$, of spinor $\psi (r)$, can be obtained from
the relation (\ref{d9a}) and it is given by%
\begin{equation*}
f_{+}(z)=-\frac{2C\sqrt{\omega _{T}}}{\varepsilon _{-}}z^{\frac{\ell +2}{2}%
}e^{-\frac{z}{2}}L_{n-1}^{\ell +\frac{3}{2}}(z).
\end{equation*}%
Normalization condition in polar coordinate is given by 
\begin{subequations}
\begin{equation}
\left\langle \psi |\psi \right\rangle =\int_{0}^{\infty }\left( \left|
f_{+}(r)\right| ^{2}+\left| f_{-}(r)\right| ^{2}\right) dr=1.  \label{d13b}
\end{equation}%
Associated Laguerre polynomials satisfy the orthogonality condition: 
\end{subequations}
\begin{equation*}
\int_{0}^{\infty }z^{\ell }e^{-z}L_{n}^{\ell }(z)L_{k}^{\ell }(z)dr=\frac{%
\Gamma \left( n+\ell +1\right) }{n!}\delta _{nk}.
\end{equation*}%
Thus, we finally obtain an expression for the normalization constant $C$:%
\begin{equation}
C=\left[ \frac{\sqrt{\omega _{T}}\varepsilon _{-}^{2}\Gamma (n+1)}{\left(
n\omega _{T}+\varepsilon _{-}^{2}\Gamma (n)\right) \Gamma \left( n+\ell +%
\frac{3}{2}\right) }\right] ^{\frac{1}{2}}.
\end{equation}

We have solved Dirac oscillator in two dimensional space which leads to a
series of interesting results. The Dirac oscillator has various physical
applications particularly in semiconductor physics \cite{villa4}.

\subsection{Coulomb Potential}

Another well known examples of the exactly solvable Dirac equation is that
relativistic Hydrogen atom. The problem can be solved exactly when the
magnetic field is zero, $A(r)=0$ and in the presence of the Coulomb
interaction \cite{villa3,lima} $v(r)=\frac{me^{2}}{4\pi \varepsilon
_{0}\hbar (\ell +1)}-\frac{\hbar (2\ell +1)}{r}.$ In this case the
superpotential is $W(r)=\frac{me^{2}}{4\pi \varepsilon _{0}\hbar ^{2}(\ell
+1)}-\frac{\ell +1}{r},$ and then the Schr\"{o}dinger-like equations (\ref%
{d10a}) and (\ref{d10b}) takes the form 
\begin{subequations}
\begin{eqnarray}
\left( -\frac{\partial ^{2}}{\partial r^{2}}+\frac{\ell (\ell +1)}{r^{2}}-%
\frac{me^{2}}{2\pi \varepsilon _{0}\hbar ^{2}r}+\left( \frac{me^{2}}{4\pi
\varepsilon _{0}\hbar ^{2}(\ell +1)}\right) ^{2}-\varepsilon ^{2}\right)
f_{-}(r) &=&0  \label{d15a} \\
\left( -\frac{\partial ^{2}}{\partial r^{2}}+\frac{(\ell +1)(\ell +2)}{r^{2}}%
-\frac{e^{2}}{r}+\left( \frac{me^{2}}{4\pi \varepsilon _{0}\hbar ^{2}(\ell
+1)}\right) ^{2}-\varepsilon ^{2}\right) f_{+}(r) &=&0  \label{d15b}
\end{eqnarray}%
Following the similar developments used in the construction of the harmonic
oscillator problem, the equation (\ref{d15a}) can be transformed in the
form: 
\end{subequations}
\begin{equation}
\left[ z\frac{\partial ^{2}}{\partial z^{2}}+\left( 2\ell +2-z\right) +n%
\right] g_{-}(z)=0  \label{ex2}
\end{equation}%
by changing the variable $r=2\pi \varepsilon _{0}\hbar ^{2}(n+\ell
+1)z/me^{2}$ and the wave function%
\begin{equation}
f_{-}(z)=Cz^{\ell +1}e^{-\frac{z}{2}}\text{ }g_{-}(z).  \label{ex3}
\end{equation}%
Natural number n and energy of the Hamiltonian satisfy the relation 
\begin{equation}
\varepsilon ^{2}=\frac{E^{2}-m^{2}c^{4}}{\hbar ^{2}c^{2}}=\left( \frac{me^{2}%
}{4\pi \varepsilon _{0}\hbar ^{2}(\ell +1)}\right) ^{2}-\left( \frac{me^{2}}{%
4\pi \varepsilon _{0}\hbar ^{2}(n+\ell +1)}\right) ^{2}.
\end{equation}%
Solution of the (\ref{ex2}) leads to the following expression for the wave
function 
\begin{subequations}
\begin{equation}
f_{-}(z)=Cz^{\ell +1}e^{-\frac{z}{2}}L_{n}^{2\ell +1}(z)  \label{d16b}
\end{equation}%
and from the relation (\ref{d9a}) we obtain 
\end{subequations}
\begin{equation}
f_{+}(z)=\frac{-C}{\varepsilon _{-}}z^{^{\ell +1}}e^{-\frac{z}{2}}\left[
2(\ell +1)L_{n-1}^{2\ell +2}(z)+(\ell +1-\frac{me^{2}}{2\pi \varepsilon
_{0}\hbar ^{2}}(2\ell +1))L_{n}^{2\ell +1}(z)\right] .
\end{equation}%
Using the identity%
\begin{equation}
\int_{0}^{\infty }z^{\ell +1}e^{-z}L_{n}^{\ell }(z)L_{n}^{\ell }(z)dr=\left(
2n+\ell +1\right) ^{\ell +2}\frac{\Gamma \left( n+\ell +1\right) }{n!},
\end{equation}%
after some straight forward calculation we obtain the normalization constant%
\begin{equation}
C=\left[ \frac{2\pi \varepsilon _{0}\hbar ^{2}}{me^{2}}\left( \frac{4(\ell
+1)^{2}}{\varepsilon _{-}^{2}\Gamma (n)}+(2n+2\ell +2)^{2\ell +3}K\right)
\Gamma (n+2\ell +2)\right] ^{-\frac{1}{2}}
\end{equation}%
where $K$ is given by%
\begin{equation}
K=\left( 1+\frac{1}{\varepsilon _{-}^{2}n!}\left( \ell +1-\frac{me^{2}}{2\pi
\varepsilon _{0}\hbar ^{2}}(2\ell +1)\right) ^{2}\right)
\end{equation}

Due to the recent interest in the 2D field theory in the condensed matter
physics, the 2D Coulomb potential is physically relevant and the results
obtained in 2D exhibit some new features \cite{ma,castro}.

\subsection{Morse Potential}

The Morse oscillator is exactly solvable quantum mechanical problem and it
is used to model the interaction of the atoms in the diatomic molecules. In
order to obtain its relativistic form we apply the standard procedure. The
choices of the parameters $v(r)=-\frac{\hbar \ell }{r}-\hbar ae^{-\alpha
r}+\hbar b,A(r)=0$ and $W(r)=b-ae^{-\alpha r},$ provides the following
potential 
\begin{subequations}
\begin{eqnarray}
\left( -\frac{\partial ^{2}}{\partial r^{2}}+a^{2}e^{-2\alpha r}-a\left(
\alpha +2b\right) e^{-\alpha r}+b^{2}-\varepsilon ^{2}\right) f_{-}(r) &=&0
\label{d17a} \\
\left( -\frac{\partial ^{2}}{\partial r^{2}}+a^{2}e^{-2\alpha r}+a\left(
\alpha -2b\right) e^{-\alpha r}+b^{2}-\varepsilon ^{2}\right) f_{+}(r) &=&0.
\label{d17b}
\end{eqnarray}%
In order to solve (\ref{d17a}) we change the variable $e^{-\alpha r}=\frac{%
\alpha }{2a}z$ and then introduce the wave function 
\end{subequations}
\begin{equation}
f_{\_}(r)=z^{\frac{b}{\alpha }-n}e^{-\frac{z}{2}}\text{ }g_{-}(z)
\end{equation}%
then we obtain%
\begin{equation}
\left[ z\frac{\partial ^{2}}{\partial z^{2}}+\left( \frac{2b}{\alpha }%
-2n+1-z\right) +n\right] g_{-}(z)=0.  \label{ex5}
\end{equation}%
Energy expression for the Morse oscillator is given by 
\begin{equation*}
\varepsilon ^{2}=\alpha n(\alpha n-2b).
\end{equation*}%
The wave functions can be obtained from the equations (\ref{ex5}) and (\ref%
{d9a}) and are given by 
\begin{subequations}
\begin{eqnarray}
f_{\_}(r) &=&Cz^{\frac{b}{\alpha }-n}e^{-\frac{z}{2}}\text{ }L_{n}^{\frac{2b%
}{\alpha }-2n}(z)  \label{d19a} \\
f_{+}(z) &=&\frac{C\alpha }{\varepsilon _{-}}z^{\frac{b}{\alpha }-n}e^{-%
\frac{z}{2}}\left[ nL_{n}^{\frac{2b}{\alpha }-2n}(z)+zL_{n-1}^{\frac{2b}{%
\alpha }-2n+1}(z)\right]  \label{d19b}
\end{eqnarray}%
Normalization condition yields the following expression for the
normalization constant 
\end{subequations}
\begin{equation}
C=\left[ \frac{\varepsilon _{-}^{2}\Gamma (n)\Gamma (n-1)}{\Gamma \left( 
\frac{2b}{\alpha }+1-n\right) \left( \varepsilon _{-}^{2}\Gamma (n)+2\alpha
^{2}\Gamma (n+1)\right) }\right] ^{-\frac{1}{2}}
\end{equation}

The potentials we have derived here is significant from both physical and
mathematical points of view. For instance the relativistic quark model
requires the solution of the dirac equation containing single quark
potential. The potential behaving like $r$ at a large distances and $1/r$ at
a short distances has been recently treated by Muci \cite{micu1}. We mention
that the Morse oscillator potential can also be used as a large distance
potential in some various physical systems. Consequently we have constructed
relativistic version of the three well known potentials whose eigenfunctions
are associated to the confluent hypergeometric functions. In the following
section we construct QES potentials.

\section{QES Potentials}

The formulation given in section 2 is also useful to construct the
relativistic version of the QES potentials. In this section we construct
three of them. At this stage we mention that the underlying idea behind the
quasi-exact solvability is the existence of a hidden algebraic structure.
Our task is now to demonstrate by appropriate choices of relativistic
potential we can construct QES Dirac equations. Our examples includes
anharmonic oscillator potential, radial sextic oscillator potential and
perturbed Coulomb potential.

\subsection{Anharmonic Oscillator potential}

The anharmonic oscillator potential have been widely used in many physical
and chemical applications. In order to construct anharmonic oscillator
potential we introduce 
\begin{equation}
v(r)=-\frac{\hbar \ell }{r}+\hbar \omega r+\hbar br^{2}+\hbar a;\ A(r)=\hbar
Br;\ W(r)=a+\omega _{T}r+cr^{2}.
\end{equation}%
With these choices we obtain the following Schr\"{o}dinger-like equations 
\begin{subequations}
\begin{eqnarray}
\left( -\frac{\partial ^{2}}{\partial r^{2}}+V_{+}+\omega
_{T}+a^{2}+\varepsilon ^{2}\right) f_{+}(r) &=&0  \label{d20a} \\
\left( -\frac{\partial ^{2}}{\partial r^{2}}+V_{-}-\omega
_{T}+a^{2}+\varepsilon ^{2}\right) f_{-}(r) &=&0  \label{d20b}
\end{eqnarray}%
where $V_{\pm }$ are given by 
\end{subequations}
\begin{equation}
V_{\pm }=2(a\omega _{T}\pm b)r+(2ab+\omega _{T}^{2})r^{2}+2b\omega
_{T}r^{3}+b^{2}r^{4}  \label{d21}
\end{equation}%
and its groundstate wave function is given by%
\begin{equation}
f_{-}^{(0)}(r)=\exp \left( -\frac{1}{3}br^{3}-\frac{1}{2}\omega
_{T}r^{2}-ar\right) ;  \label{d22}
\end{equation}%
A part of the spectrum of the anharmonic oscillator potential can be
obtained in the framework of the QES problem or its approximate solution can
be obtained by using perturbation theory.

\subsection{Sextic Oscillator potential}

Another well known QES potential is the sextic oscillator potential. Its
relativistic form can be obtained by the following choices of functions: 
\begin{equation}
v(r)=-\frac{2\hbar \ell }{r}+\hbar \omega r+\hbar br^{3};A(r)=\hbar Br;W(r)=-%
\frac{\ell }{r}+\omega _{T}r+br^{3}.
\end{equation}%
In this case we obtain the following equations: 
\begin{subequations}
\begin{eqnarray}
\left( -\frac{\partial ^{2}}{\partial r^{2}}+\frac{\ell (\ell +1)}{r^{2}}%
+V_{+}+\varepsilon ^{2}\right) f_{+}(r) &=&0  \label{d23a} \\
\left( -\frac{\partial ^{2}}{\partial r^{2}}+\frac{\ell (\ell -1)}{r^{2}}%
+V_{-}+\varepsilon ^{2}\right) f_{-}(r) &=&0  \label{d23b}
\end{eqnarray}%
where 
\end{subequations}
\begin{equation}
V_{\pm }=(\omega _{T}^{2}\pm b(2\ell \mp 3))r^{2}+2b\omega
_{T}r^{4}+b^{2}r^{6}-\omega _{T}(2\ell \mp 1)  \label{d24}
\end{equation}%
Its groundstate wave function is given by%
\begin{equation}
f_{-}^{(0)}(r)=r^{\ell }\exp \left( -\frac{1}{2}\omega _{T}r^{2}-\frac{1}{4}%
br^{4}\right)   \label{d25}
\end{equation}

\subsection{Deformed Coulomb potential}

Our last example is the deformed Coulomb potential which can be obtained by
introducing%
\begin{equation}
v(r)=\frac{e^{2}\hbar }{2(\ell +1)}-\frac{\hbar (2\ell +1)}{r};\ A(r)=\hbar
Br;\ W(r)=\frac{e^{2}}{2(\ell +1)}-\frac{\ell +1}{r}.
\end{equation}%
In this case the equations take the forms

\begin{subequations}
\begin{eqnarray}
\left( -\frac{\partial ^{2}}{\partial r^{2}}-\frac{(\ell +1)(\ell +2)}{r^{2}}%
+V_{+}-\omega _{T}(2\ell +1)+\frac{e^{4}}{4(\ell +1)^{2}}+\varepsilon
^{2}\right) f_{+}(r) &=&0  \label{d26a} \\
\left( -\frac{\partial ^{2}}{\partial r^{2}}-\frac{\ell (\ell +1)}{r^{2}}%
+V_{-}-\omega _{T}(2\ell +3)+\frac{e^{4}}{4(\ell +1)^{2}}+\varepsilon
^{2}\right) f_{-}(r) &=&0  \label{d26b}
\end{eqnarray}%
\end{subequations}
\begin{equation}
V_{\pm }=-\frac{e^{2}}{r}+\omega _{T}^{2}r^{2}+\frac{e^{2}\omega _{T}}{\ell
+1}r+\frac{e^{4}}{4(\ell +1)^{2}}
\end{equation}%
Its groundstate eigenfunction are given by%
\begin{equation}
f_{-}^{(0)}(r)=r^{\ell +1}\exp \left( -\frac{\omega _{T}}{2}r^{2}-\frac{%
e^{2}r}{2(\ell +1)}\right) .
\end{equation}%
The non-relativistic two dimensional Hamiltonian described Coulomb
interaction between charged particles, i. e. the interaction between
conduction electron and donor impurity center when a constant magnetic field
is applied perpendicular to the plane of motion has been discussed in the
literature\cite{villa3,taut,mustafa,koc}. We have obtained relativistic
interaction and it can be solved quasi-exactly. Recently its spectrum and
wavefunction has been computed numerically \cite{villa1}.

\section{Conclusion}

We have obtained analytical solutions of the $(2+1)$- dimensional Dirac
equation for a set of potentials in two dimensions with the hope that they
could be useful in the low dimensional field theory and condensed matter
physics. The potentials for Dirac equation have been obtained by extending
the notion of the Dirac oscillator. In a similar manner one can construct
the exactly and QES Dirac equation including hyperbolic and trigonometric
potentials. It has been shown that the $(2+1)$-dimensional Dirac equation
can be transformed in the form of a Schr\"{o}dinger-like equation and for
all exactly and QES Schr\"{o}dinger equations one can find potentials of the
Dirac equation which are also exactly solvable or QES.

Therefore, besides its importance as a new treatment of the construction of
the various potentials for Dirac equation, the potentials obtained here
might be relevant to model some physical problems. Before ending this work a
remark is in order. It is expected that the models presented here to
construct Dirac equation including various potentials may provide a good
starting point for the study of more realistic models for the low
dimensional of field theory and condensed matter physics.


\begin{thebibliography}{9}

\bibitem{villa1} V. M. Villalba and R. Pino, \textit{Mod.Phys.Lett.B }%
\textbf{17} (2003) 1331.

\bibitem{chak} A. M. J. Schakel, \textit{Phys. Rev. D}. \textbf{43} (1991)
1428.

\bibitem{villa2} V. M. Villalba, \textit{Phys. Rev. A }\textbf{49} (1994)
586.

\bibitem{schak} A. M. J. Schakel and G. W. Semenoff, \textit{Phys. Rev. Lett}%
.\textbf{\ 66} (1991) 2653.

\bibitem{naagu} A. Neagu and A. M. J. Schakel, \textit{Phys. Rev. D.} 
\textbf{48} (1993) 1785.

\bibitem{ma} S. H. Dong and Z. Q. Ma, \textit{Phys. Lett. A} \textbf{312}
(2003) 78.

\bibitem{lin} Q. G. Lin, \textit{Phys. Lett. A} \textbf{260} (1999) 17.

\bibitem{vaid} A. N. Vaidya and L. E. S. Souza, \textit{Phys. Lett. A} 
\textbf{293} (2002) 129.

\bibitem{khal} V. R. Khalilov and C. L. Ho,\textit{\ Mod. Phys. Lett. A} 
\textbf{13} (1998) 615.

\bibitem{wil} F. Wilczek, \textit{Phys. Rev. Lett.} \textbf{48} (1982) 1144.

\bibitem{fesh} H. Feshbach and F. Villars, \textit{Rev. Mod. Phys.} \textbf{%
30} (1958) 24.

\bibitem{merad} M. Merad, L. Chetouani and A. Bounames, \textit{Phys. Lett. A%
} \textbf{267} (2000) 225.

\bibitem{robson1} B. A. Robson and D. S. Staudte, \textit{J. Phys. A: Math.
Gen.} \textbf{29 }(1996) 157.

\bibitem{robson2} D. S. Staudte, \textit{J. Phys. A: Math. Gen.} \textbf{29 }%
(1996) 169.

\bibitem{ito} D. It\^{o}, K. Mori and E. Carriere, \textit{Nuovo Cimento A} 
\textbf{51} (1967) 1119.

\bibitem{mosh1} M. Moshinsky and A. Szczepaniak, \textit{J. Phys. A: Math.
Gen.} \textbf{22} (1989) L817.

\bibitem{moreno} M. Moreno and A. Zentella, \textit{J. Phys. A: Math. Gen.} 
\textbf{22} (1989) L821.

\bibitem{mosh2} M. Moshinsky, and Y. F. Smitnov, The Harmonic Oscillator in
Modern Physics, Harwood Academic Publishers, Amsterdam, 1996.

\bibitem{ferk} N. Ferous and A. Bounames, \textit{Phys. Lett. A} \textbf{325}
(2004) 21.

\bibitem{ros} P. Rojmez, R. and Arvieu, \textit{J. Phys. A: Math. Gen.} 
\textbf{32} (1999) 5367.

\bibitem{ho} C. L. Ho and P. Roy, \textit{Ann. Phys.} \textbf{312} (2004)
161.

\bibitem{mart} R. Mart\'{\i}nez, M. Moreno and A. Zentella, \textit{Rev.
Mex. F\'{\i}s.} \textbf{36} (1990) S176.

\bibitem{levai} G. Levai and A. Del Sol Mesa, \textit{J. Phys. A: Math. Gen.}
\textbf{29} (1996) 2827.

\bibitem{alhaidari} A. D. Alhaidari, \textit{J. Phys. A: Mat. Gen.} \textbf{%
34} (2001) 9827.

\bibitem{turb} A. V. Turbiner and A. G. Ushveridze, \textit{Phys Lett. A} 
\textbf{126}, (1987) 81.

\bibitem{bender1} C. M. Bender and M. Moshe, \textit{Phys. Rev. A} \textbf{55%
} (1997) 2625.

\bibitem{bender2} C. M. Bender and S. Boettcher, \textit{J. Phys. A: Math.
Gen.} \textbf{31} (1998) L273.

\bibitem{villa3} V. M. Villalba and R. Pino, \textit{Phys. Lett. A} \textbf{%
238}, (1998) 49.

\bibitem{villa4} V. M. Villalba and A. R. Maggiola, \textit{Eur. Phys. J. B} 
\textbf{22}, (2001) 31.

\bibitem{micu1} L. Micu, Mod. \textit{Phys. Lett. A} \textbf{17}, (2003)
2895.

\bibitem{taut} M. Taut, \textit{J. Phys. A: Math. Gen.} \textbf{28} (1995)
2081.

\bibitem{mustafa} O. Mustafa, \textit{J. Phys. : Condens. Matter} \textbf{5}%
, (1993), 1327.

\bibitem{koc} R. Ko\c{c}, H. T\"{u}t\"{u}nc\"{u}ler and E. Ol\u{g}ar, 
\textit{J. Kor. Phys. Soc.} To be published.

\bibitem{lima} R. L. Rodrigues, \textit{Phys. Lett. A} \textbf{326} (2004)
42.

\bibitem{castro} A. S. de Castro, \textit{Ann. Phys.}, \textbf{311} (2004)
170.

\bibitem{roy} C. L. Ho and P. Roy, \textit{Ann. Phys.,} \textbf{312} (2004)
161.
\end{thebibliography}
\end{document}